# Enhancing Brain Tumor Classification Using TrAdaBoost and Multi-Classifier Deep Learning Approaches

Mahin Mohammadi[1], Saman Jamshidi[2]

[1]Department of Health Information Management, School of Health Management and Information Sciences, Iran University of Medical Sciences, Tehran, Iran
[2] Department of Management Science and Technology, Amirkabir University of Technology, (Tehran Polytechnic), Tehran, Iran

## Abstract

Brain tumors pose a serious health threat due to their rapid growth and potential for metastasis. While medical imaging has advanced significantly, accurately identifying and characterizing these tumors remains a challenge. This study addresses this challenge by leveraging the innovative TrAdaBoost methodology to enhance the Brain Tumor Segmentation (BraTS2020) dataset, aiming to improve the efficiency and accuracy of brain tumor classification.

Our approach combines state-of-the-art deep learning algorithms, including the Vision Transformer (ViT), Capsule Neural Network (CapsNet), and convolutional neural networks (CNNs) such as ResNet-152 and VGG16. By integrating these models within a multi-classifier framework, we harness the strengths of each approach to achieve more robust and reliable tumor classification. A novel decision template is employed to synergistically combine outputs from different algorithms, further enhancing classification accuracy.

To augment the training process, we incorporate a secondary dataset, "Brain Tumor MRI Dataset," as a source domain, providing additional data for model training and improving generalization capabilities. Our findings demonstrate a high accuracy rate in classifying tumor versus non-tumor images, signifying the effectiveness of our approach in the medical imaging domain. This study highlights the potential of advanced machine learning techniques to contribute significantly to the early and accurate diagnosis of brain tumors, ultimately improving patient outcomes.

# 1.Introduction

Medical image processing represents a cutting-edge technological advancement in contemporary research. Noninvasive techniques can be effectively utilized through various medical imaging modalities. Key noninvasive imaging technologies include Positron Emission Tomography (PET), Computed Tomography (CT), Ultrasound, Single Photon Emission Computed Tomography (SPECT), Magnetic Resonance Imaging (MRI), and X-ray. Among these, MRI is known to deliver superior results compared to computed tomography in the realm of medical diagnostics. It enhances the contrast between different soft tissues in the human body, thereby improving diagnostic accuracy[1].

According to the World Health Organization (WHO) and the American Brain Tumor Association (ABTA), the most widely accepted classification scheme for brain tumors uses a grading system that ranges from grade I to grade IV. This system effectively distinguishes between different types of benign growths and malignant tumors[2].

Brain tumors are classified into two main types: benign (non-cancerous) and malignant (cancerous). Malignant tumors have the potential to spread rapidly to other tissues within the brain, which can significantly worsen the patient's condition[3]. Normally, old or damaged cells are replaced by new, healthy cells. However, if these damaged or aged cells are not eliminated during the generation of new cells, it can lead to complications. This excess production of cells often results in the formation of a mass of tissue, commonly referred to as a tumor. Detecting brain tumors is particularly challenging due to factors such as the size, shape, location, and type of tumor. Early diagnosis of brain tumors is often difficult, as accurate measurements of tumor size and resolution are not easily attainable[4].

Magnetic Resonance Imaging (MRI) is the most accurate imaging modality for diagnosing various types of brain cancers. Unlike traditional X-rays, MRI scans utilize powerful magnetic fields and radio waves to create detailed images of the brain. As noted by researchers Josefina Perlo, Christoph Mülder, Ernesto Danieli, Christian Hopmann, Bernhard Blümich, and Federico Casanova, MRI produces visual "slices" of the brain that can be compiled to create a three-dimensional representation of tumors[5].

To enhance the clarity of these scans, contrast agents may be employed. An MRI scan operates by using specific components of its intense magnetic field to generate radio frequency pulses, resulting in comprehensive images of organs, soft tissues, bones, and other internal structures of the human body[6, 7].

This technique is invaluable for providing extensive information regarding the architecture of brain tumors, enabling accurate diagnosis, effective treatment planning, and monitoring of disease progression. Advances in MRI technology have refined the ability to measure changes within and around primary and metastatic brain tumors, including edema, volumetric alterations, and distinct anatomical features. furthermore, MRI can produce computerized representations of tissue characteristics across various planes, thereby facilitating the comprehensive detection and assessment of brain masses. Collectively, these capabilities underscore MRI's critical role in the clinical evaluation of brain tumors[6, 7].

With the advances in machine learning and computational techniques in recent years, the potential and the need of developing computerized methods to assist radiologists in image analysis and diagnosis has been recognized as an important area of research and development in medical imaging[8]

The rapid advancement and adoption of big data and AI technologies have revolutionized problem-solving processes by enabling real-time, data-driven predictions of various diseases, comprehensive evaluations of treatment options, and the automated execution of complex tasks on a large scale[9].

Deep learning methods excel particularly well when trained on substantial datasets. Significant gains in computer vision have further propelled the application of deep learning in medical image analysis, encompassing tasks such as segmentation, object detection, classification, prognosis prediction, and microscopic imaging analysis. Among the array of computational tools available, deep learning-based applications are emerging as the gold standard, enhancing accuracy and unlocking new possibilities in the field of medical image analysis[9].

The focal point of our study lies in the innovative application of the groundbreaking TrAdaBoost [10] methodology, strategically implemented to enhance the Brain Tumor Segmentation (BraTS2020) dataset[11] under examination. This strategic augmentation not only reduces learning costs but also elevates learning efficiency to unprecedented levels.

Within the architectural blueprint of our pioneering framework, a multitude of cutting-edge algorithms takes center stage. Incorporating essential mechanisms such as the Attention Mechanism, Vision Transformer (ViT)[12], Capsule Neural Network (CapsNet)[13], and Decision Templates significantly boosts the efficiency of the segmentation process, leading to more accurate and robust results. The integration of state-of-the-art algorithms like the Attention Mechanism, ViT, CapsNets, and Decision Templates refines feature relevance, semantic relationships, and decision-making guidance during the segmentation of intrinsically heterogeneous brain tumors.

Additionally, utilizing ResNet-152[14] and VGG 16[15] models to capture image features enhances the depth and breadth of our proposed model's capabilities. Our model's complexities are clearly illustrated in the visual representation below, which provides a comprehensive overview of the interconnected components and their synergistic integration that underpins our approach to brain tumor diagnosis, specifically focusing on the evaluation of state-of-the-art methods for segmenting gliomas in multimodal magnetic resonance imaging (MRI) scans within the BraTS2020 framework.

## 2. Related Works

Recently, deep learning has achieved remarkable advancements in various medical image segmentation tasks, particularly when ample training data is available. The multi-modal Brain Tumor Segmentation (BraTS) challenge has provided a substantial dataset of pre-operative multi-modal MRI images along with their corresponding manual annotations for brain tumors [16, 17]. This rich dataset has propelled convolutional neural networks (CNNs) to dominate the field of fully automated brain tumor segmentation [18, 19].

CNN methods can be categorized into 2D slice-based or 3D patch-based approaches. In 2D CNN methods, a 3D volume is partitioned into multiple 2D slices, and brain tumors are predicted independently for each slice[18, 20]. For example, Caver et al. employed three separate 2D UNets to segment whole tumors (WT), tumor cores (TC), and enhancing tumors (ET) in a slice-by-slice manner [21]. Similarly, Choudhury et al. trained three multi-class segmentation models on axial, coronal, and sagittal slices, utilizing majority voting to generate final predictions [22]. McKinley et al. introduced a novel DeepSCAN architecture that integrates a pooling-free DenseNet within a UNet framework, applying their model across three planes and using majority voting for final predictions. This multi-view fusion technique of 2D CNN results has also been widely applied to other tasks, referred to as 2.5D CNN[23, 24].

Additionally, other types of deep learning networks have been explored for brain tumor segmentation, including generative adversarial networks (GAN), transformers, and capsule neural networks. For instance, Nema et al. developed a network architecture called the residual cyclic unpaired GAN (RescueNet), which incorporates residual and mirroring principles[25]. Wang et al. integrated a transformer into a 3D CNN for brain tumor segmentation, proposing a novel network known as TransBTS, built on an encoder-decoder structure [26]. Aziz et al. optimized a capsule neural network-based architecture, named SegCaps, to achieve accurate glioma segmentation from MRI images[27] .

The BraTS 2021 challenge employs a comprehensive benchmarking approach to evaluate brain glioma segmentation algorithms using meticulously curated multi-parametric Magnetic Resonance Imaging (mpMRI) data from 2,040 patients. This edition focuses on two primary tasks: (1) the segmentation of histologically distinct brain tumor sub-regions and (2) the classification of the tumor's O[6]-methylguanine-DNA methyltransferase (MGMT) promoter methylation status. Performance assessments for participating algorithms are conducted through the Sage Bionetworks Synapse platform for Task 1 and Kaggle for Task 2. Monetary awards totaling $60,000 will be distributed to the top-ranked participants, underscoring the competitive and collaborative nature of this challenge[28].

In another investigation, the segmentation of gliomas—a serious brain tumor impacting surgical planning and progression evaluation—was approached by leveraging prior knowledge, training strategies, and ensemble methods. The study delineated three partially-overlapping regions of interest: whole tumor (WT), tumor core (TC), and enhancing tumor (ET). The proposed segmentation pipeline involved three UNet architectures with varying inputs to enhance segmentation performance. The first UNet utilized 3D patches from multi-modal MR images, while the second integrated brain parcellation as an additional input. The third UNet combined 2D slices of multi-modal MR images with brain parcellation and probability maps for WT, TC, and ET generated by the second UNet. This sequential process united WT segmentation from the third UNet with the fused TC and ET segmentations from the first and second UNets, constituting the complete tumor segmentation. A post-processing strategy was implemented to correct false-positive ET segmentations by labeling small ET areas as non-enhancing tumors[29].

Jaspin and Suganthi [30]conducted a study on brain tumor segmentation and morphological edge detection in MR images, employing regional growth (RG) techniques. Their performance evaluation utilized the fuzzy C-means (FCM) approach. The study integrated three methodologies: RG, region of interest (ROI), and morphological operations. The proposed solution included preprocessing to eliminate noise, area growth based on FCM, and edge detection via morphological operations to enhance the images. Following these morphological processes, which included erosion and dilation on the tissue samples, the FCM technique was subsequently used for tumor segmentation.

Singh et al. [31]introduced an innovative method for brain tumor detection that utilized a fully connected pyramid pooling network (FCPPN) for tumor segmentation. This approach aimed to accurately identify the location and type of tumor for classification. The classification was performed using a multi-tier convolutional neural network with channel preference (MTCNNCP). Once classified, predicting patient survival can be complex; to address this, the authors developed a prediction model called Multi-tier Zernike (MTZR), which operates on synthetic choices. The severity of the tumor is assessed through the calculation of geometric distance.

To enhance image quality, the same authors [32]proposed a novel adaptive diffusivity function defined by partial differential equations for denoising. This diffusivity function leverages gradient, Laplacian, and adaptive thresholding techniques to improve brain tumor images while preserving essential details. The refined images are then processed using an improved multi-kernel fuzzy c-means (MKFCM) algorithm for accurate image segmentation, effectively distinguishing between tumor and normal tissue.

## 3. Methodology

In this section, we present the methodologies and framework utilized for brain tumor diagnosis.

### 3.1 Architecture of the proposed model for brain tumor diagnosis

Figure 1 provides an overview of our model, which employs several robust algorithms to enhance performance. As illustrated, our model integrates three classifiers. The first two classifiers utilize pre-trained VGG16 and ResNet-50 models to extract image features. Additionally, both classifiers implement the TrAdaBoost algorithm to enhance data quality and improve overall results.

The first classifier leverages the capabilities of Vision Transformer (ViT) alongside Multi-Head Self-Attention, while the second classifier utilizes Capsule Neural Network (CapsNet) in conjunction with Multi-Head Self-Attention.

The third classifier operates solely using convolutional layers and fully connected layers. To combine the outputs of these three classifiers, we apply a decision template algorithm, ensuring a cohesive integration of results for improved accuracy.

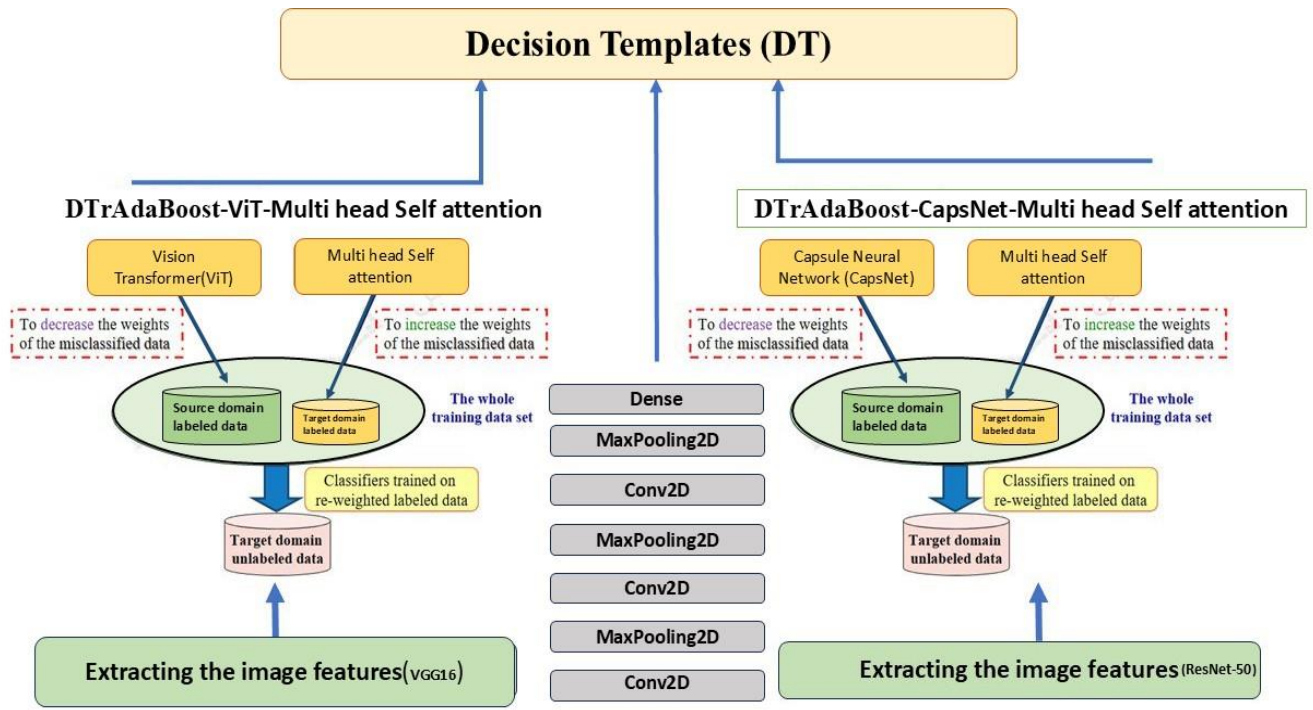

Figure 1 The architecture of the model provided to diagnose brain tumor

## 3.2 Transfer learning

Like artificial intelligence (AI) and machine learning (ML) in general, the concept of transfer learning has evolved over decades. From the very beginning of AI research, researchers have recognized the ability to transfer knowledge as a fundamental aspect of intelligent behavior [33].

Early explorations of transfer learning were conducted under various names within the field of AI, including "learning by analogy", "case-based reasoning (CBR)", "knowledge reuse and re-engineering", "lifelong machine learning", "never-ending learning", "fine-tuning", and "domain adaptation." These diverse terminologies reflect the breadth of approaches taken to address the fundamental challenge of leveraging past knowledge to facilitate learning in new situations[33].

Transfer learning (TL) draws inspiration from cognitive research, which posits that humans learn new tasks more effectively by leveraging existing knowledge from similar tasks. This idea translates to artificial intelligence, where TL allows models to improve performance on a target task by transferring knowledge from a related source task[34].

Formally, Pan and Yang define TL using the concepts of domains and tasks. A domain encompasses a feature space X and its associated probability distribution P(X), while a task involves a label space Y and a predictive function f(·) that maps features to labels. Transfer learning aims to improve the prediction function fT(·) for a target task T T in a target domain D T by utilizing knowledge acquired from a source domain D S and source task T S[34].

### 3.2.1. Extracting image features using VGG 16 and ResNets 50 models

Transfer learning (Sertolli, Ren, Schuller & Cummins, 2021) is carefully examined in this phase. It proves particularly valuable when training data is limited, as it can achieve performance comparable to that of healthcare experts in disease diagnosis, as demonstrated in studies [35] and[36]. This approach can also be applied to the case of uveal melanoma (UM).In the context of disease diagnosis research, there are proposals to utilize transfer learning networks for identifying brain tumor[37]. Specifically, a VGG (VIN)-inspired network is suggested as a backbone, which serves to take images as input and extract feature maps for further analysis[38].

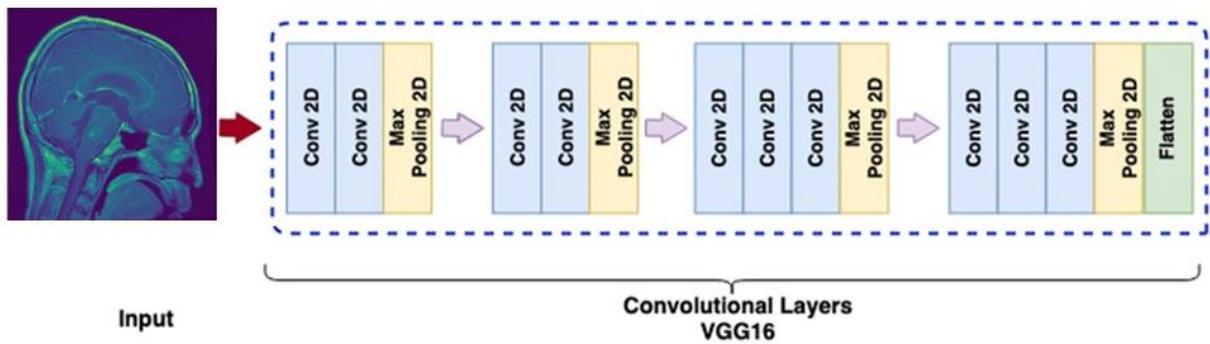

Figure 2 Extracting features of brain tumor MRI images using VGG 16 model

Residual Neural Networks (ResNets) have revolutionized deep learning by enabling the training of incredibly deep networks, containing hundreds or even thousands of layers, while still achieving exceptional performance. This breakthrough in representational power has significantly improved numerous computer vision applications beyond image classification, including object detection and facial recognition[39].

The universal approximation theorem states that a single-layer feedforward network can approximate any function given sufficient capacity. However, this theoretical capability often translates to massive networks prone to overfitting. Furthermore, the well-known vanishing gradient problem hinders the training of deep networks. As gradients are backpropagated through numerous layers, repeated multiplication can cause them to shrink to negligible values, effectively hindering learning in earlier layers. Consequently, as the network deepens, performance can plateau or even degrade rapidly[39]. ResNets address this challenge by introducing "skip connections," allowing information to bypass multiple layers. This bypass mechanism prevents vanishing gradients and facilitates the flow of information throughout the network, enabling effective training of extremely deep architectures[39].

Like the VGG-16 model, ResNets have also been widely adopted as backbones for extracting image features in various computer vision tasks. Their robust feature extraction capabilities make them a powerful tool for a wide range of applications.

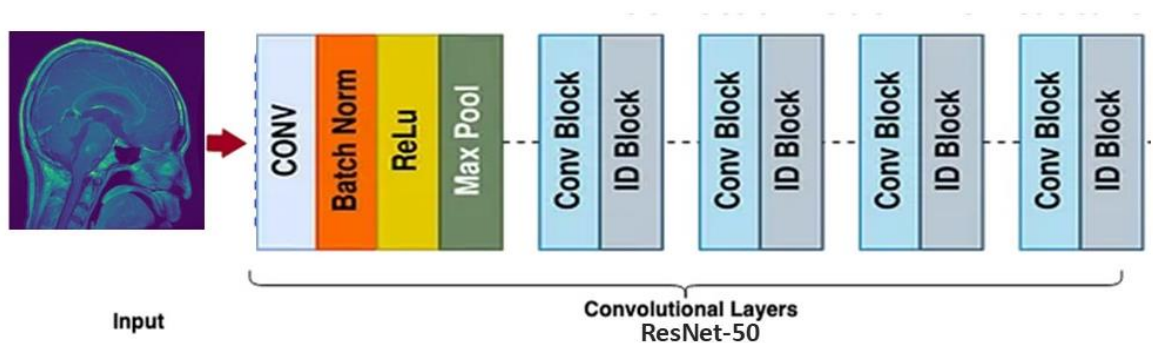

Figure 3 Extracting brain tumor MRI image features using ResNets 50 model

### 3.2.2. Transfer AdaBoost algorithm

Transfer AdaBoost algorithm, also called TrAdaBoost, is a classic transfer learning algorithm which is proposed by Dai et al.[40]. TrAdaBoost assumes that the source and target domain data use exactly the same set of features and labels, but the distributions of the data in the two domains are different. In addition, TrAdaBoost assumes that, due to the difference in distributions between the source and the target domains, some of the source domain data may be useful in learning for the target domain but some of them may not and could even be harmful[41].

### 3.3 Capsule neural networks

Capsule neural networks (CapsNets) are a type of artificial neural network (ANN) designed to better model hierarchical relationships, drawing inspiration from the organization of biological neural systems[42].

CapsNets augment convolutional neural networks (CNNs) with "capsules," specialized structures that encode information about an object's properties, such as its pose and presence. Outputs from multiple capsules are combined to form more stable representations for higher-level capsules, making them less susceptible to variations in input data [43].Each capsule outputs a vector representing the probability of an object's presence and its pose, akin to classification with localization in traditional CNNs.

One key advantage of CapsNets is their ability to address the "Picasso problem" in image recognition, where images with correctly placed parts but incorrect spatial relationships are misclassified. [44] CapsNets exploit the observation that viewpoint changes have nonlinear effects at the pixel level but linear effects at the part/object level[44]. This allows them to effectively learn the spatial relationships between object parts, similar to inverting the rendering process of an object with multiple parts[45].

In essence, CapsNets offer a more sophisticated approach to object representation and recognition, capturing hierarchical relationships and spatial information in a manner that better aligns with biological neural systems.

### 3.4 Vision Transformer

A Vision Transformer (ViT) is a specialized transformer architecture tailored for computer vision tasks. Unlike traditional methods that convert text into tokens, a ViT divides an input image into a series of patches, serializes each patch into a vector, and then reduces its dimensionality with a single matrix multiplication. These vector embeddings are subsequently processed by a transformer encoder in the same manner as token embeddings[46].

ViTs were introduced as alternatives to Convolutional Neural Networks (CNNs) in the realm of computer vision[47]. They exhibit distinct inductive biases, varying training stability, and differences in data efficiency. Although ViTs are generally less data-efficient compared to CNNs, they possess a higher capacity for learning. Some of the most advanced computer vision models today are ViTs, including one that boasts an impressive 22 billion parameters[48].

Since their inception, numerous variants have emerged, including hybrid architectures that integrate features from both ViTs and CNNs. Vision Transformers have proven effective in a range of applications, including image recognition, image segmentation, and autonomous driving[49].

### 3.5 Multi-Head Self-Attention

The Transformer architecture, renowned for its success in natural language processing, employs a powerful mechanism known as multi-head attention. This mechanism allows the model to capture complex relationships and nuances between words in a sentence by performing multiple parallel attention calculations[50, 51].

**The Core Concept:**

At its heart, the attention module within the Transformer uses three sets of parameters: queries (Q), keys (K), and values (V). These parameters are used to calculate attention scores, which represent the relevance of each word in the input sequence to a specific target word. In multi-head attention, the Q, K, and V parameters are split into multiple "heads", typically denoted by 'N'. Each head operates independently, performing its own attention calculation[50].

**Parallel Processing for Enhanced Understanding:**

Instead of relying on a single attention head to capture all relationships, multi-head attention allows the Transformer to simultaneously focus on different aspects of the input sequence through its multiple heads[52, 53].

In the visual representation provided as 4th image, we can observe the type of Decision Fusion used. If we consider the n-dimensional input, our input set is defined as $X = [x_1, x_2, x_3, .., x_n]^T$. Also, $L = [l_1, l_2, l_3, .., l_h]$ ($h$: Number of labels) be the set of potential class labels; and $C =$

$[C_1, C_2, C_3, .., C_l]$ ($l$: Number of Classifiers) be the set of trained classifiers for decision fusion. Given the input set, the output of the classifier for each input is [53, 54]

$$C_i(X) = [c_{i,1}, c_{i,2}, c_{i,3}, .., c_{i,h}]^T. \quad (1)$$

The fused output of $l$ classifiers is produced as in $C(X) = F(C_1(X), C_2(X), ... C_l(X))$. We can represent the output of all the classifiers in a Decision Profile (DP); Decision Profile is a $l \times h$ matrix [53, 54]:

$$DP(X) = \begin{bmatrix} C_{1,1}(X) & \cdots & C_{1,j}(X) & \cdots & C_{1,h}(X) \\ \vdots & & & & \vdots \\ C_{i,1}(X) & \cdots & C_{i,j}(X) & \cdots & C_{i,h}(X) \\ \vdots & & & & \vdots \\ C_{l,1}(X) & \cdots & C_{l,j}(X) & \cdots & C_{l,h}(X) \end{bmatrix}, \quad (2)$$

The $i$-th row represented the $i$-th classifier output for each $h$ label. In addition, the fusion result is a $h$-dimensional vector that is represented by the measure layer form, which is denoted as in $DT = [A_1(X), A_2(X), A_3(X), ..., A_h(X)]^T$, where $A_i$ shows which determine the measurement value of each label [55]. For each label, we consider $N$ samples related to that label detected by the classifiers, then we get the average of these $N$ samples and we consider it as A, Simply, the value for each label is calculated as follows. We called this Decision Table (DT) where $= 1, 2, ..., h, N = 100, j = 1, 2, ..., m$, and $m$ is the train input sample

$$A_i = \frac{1}{N} \sum_1^N C_j, \quad (3)$$

$$\text{Deuc} = \sqrt{(x - y)^2}. \quad (4)$$

In this paper, we work with two labels, denoted as h = 2. After training the model, we utilize the training data to create Decision Points (DP) and Decision Tables (DT). Based on the value of H, we generate two decision tables. We then apply the Euclidean Distance formula to assign each data point to its corresponding label. For each data point, we compute a value between 0 and 1: a value closer to 0 indicates that the data point belongs to the first label, while a value closer to 1 signifies its association with the second label.

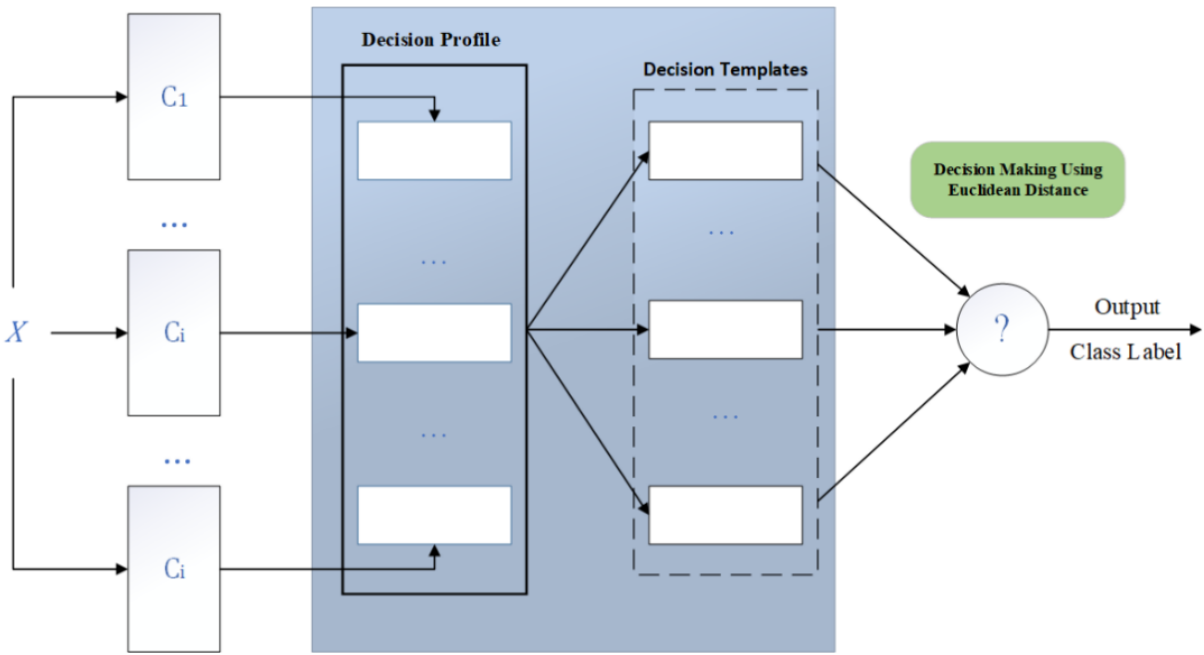

Figure 4The architecture of the applied Decision Fusion[53]

## 4.Experiments and results

### 4.1 Dataset

In this study, we utilized two datasets sourced from Kaggle. Our target domain dataset is the "Brain Tumor Segmentation (BraTS2020)" dataset, which is freely accessible at this link (https://www.kaggle.com/datasets/awsaf49/brats2020-training-data). For our source domain dataset, we selected the "Brain Tumor MRI Dataset," available at this link (https://www.kaggle.com/datasets/masoudnickparvar/brain-tumor-mri-dataset/data).
From the target dataset, we extracted 500 MRI image samples of glioma tumors and 400 notumor samples. In the source domain dataset, we selected 200 glioma tumor samples and 250 notumor samples for our analysis.

### 4.2 The results of the training and evaluation phase

The implementation environment for this study was Google Colab. We developed the models using the Python programming language along with the Keras library. To optimize our model training, we allocated 80% of the samples for training and reserved 20% for testing. Prior to inputting the images into the models, we normalized them by scaling the pixel values to a range between 0 and 1.

Figure 5 illustrates the error and accuracy metrics for the VGG16 and DTrAdaBoost-ViT Multi-Head Self-Attention model during training. The model was trained across 10 epochs, and the results indicate a clear trend: as the number of IPACs increases, both the training and validation error rates decline, while the accuracy for both datasets rises. This demonstrates the effectiveness of the model in improving performance with additional training iterations.

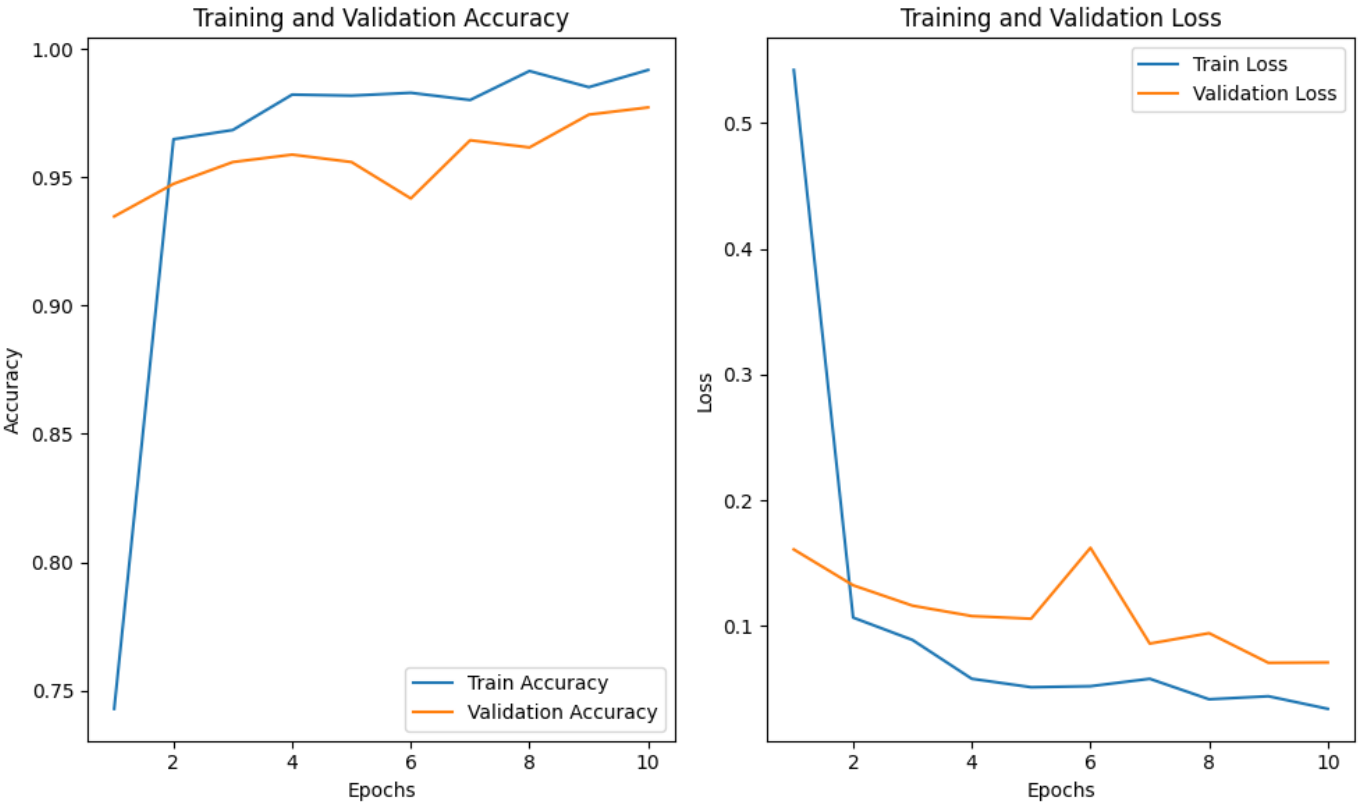

Figure 5 Accuracy and error diagrams VGG 16, DTrAdaBoost-ViT-Multi head Self attention

The training results for the ResNet-50-DTrAdaBoost-CapsNet-Multi-Head Self-Attention and CNN models are presented in Figures 6 and 7. Both classifiers were trained using 10 epochs. Similar to the previous classifier, the accuracy for these two models shows a consistent upward trend, while the error rates for both training and validation datasets exhibit a downward trajectory. This alignment in performance improvement underscores the effectiveness of the training process.

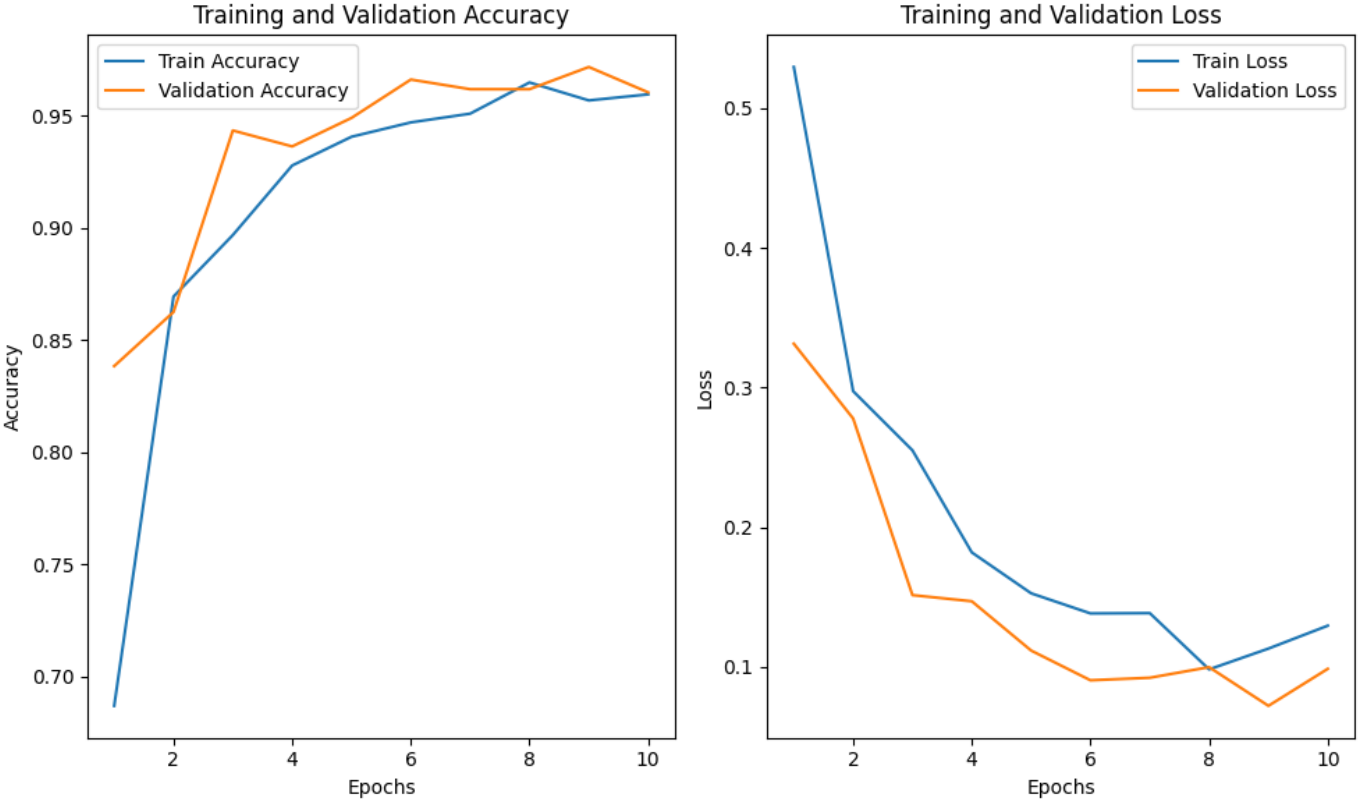

*Figure 6* Accuracy and error diagrams ResNet-50-DTrAdaBoost-CapsNet-Multi-Head Self-Attention

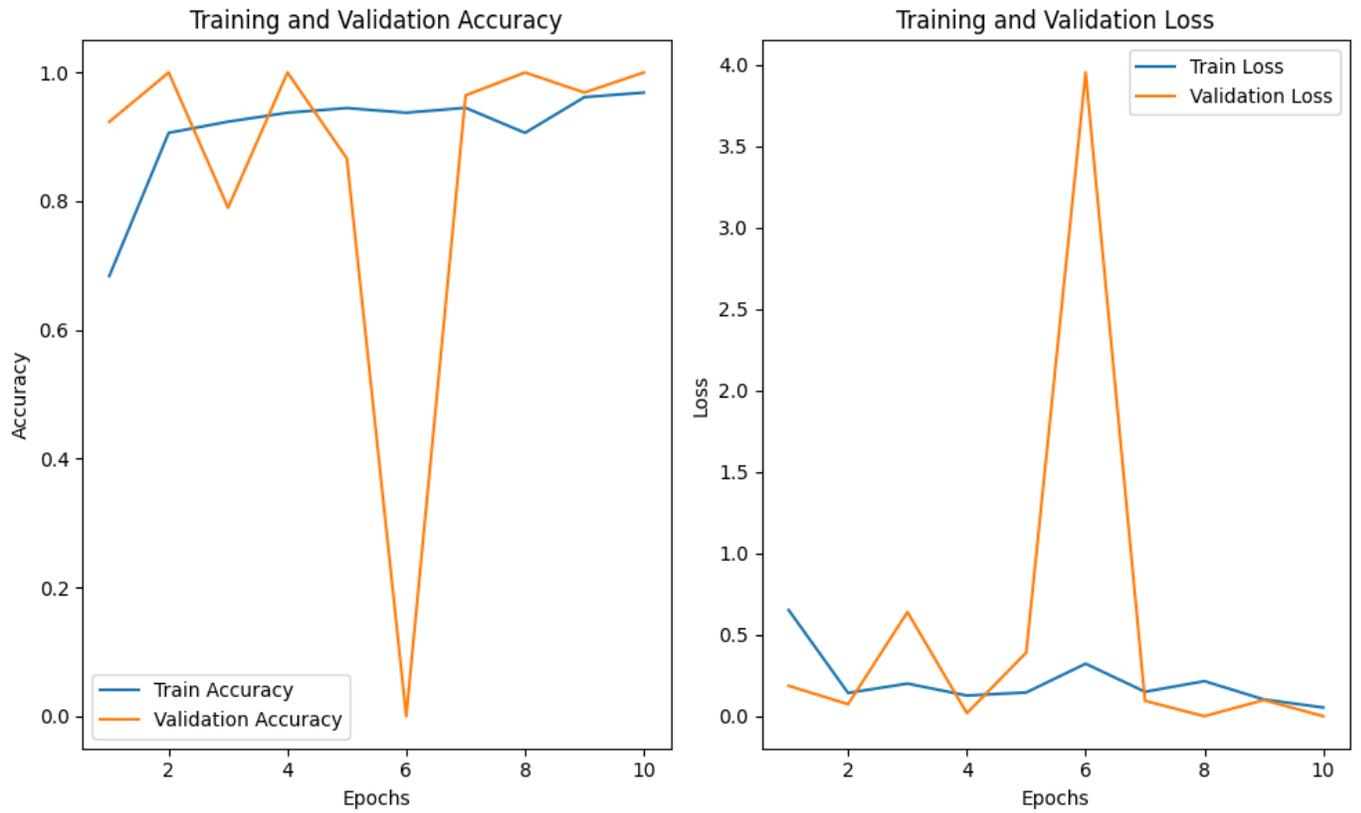

*Figure 7Accuracy and error diagrams CNN*

Figure 8 presents the confusion matrix values for our proposed method applied to the test data. From this data, we can conclude that the accuracy of our model is an impressive 99%.

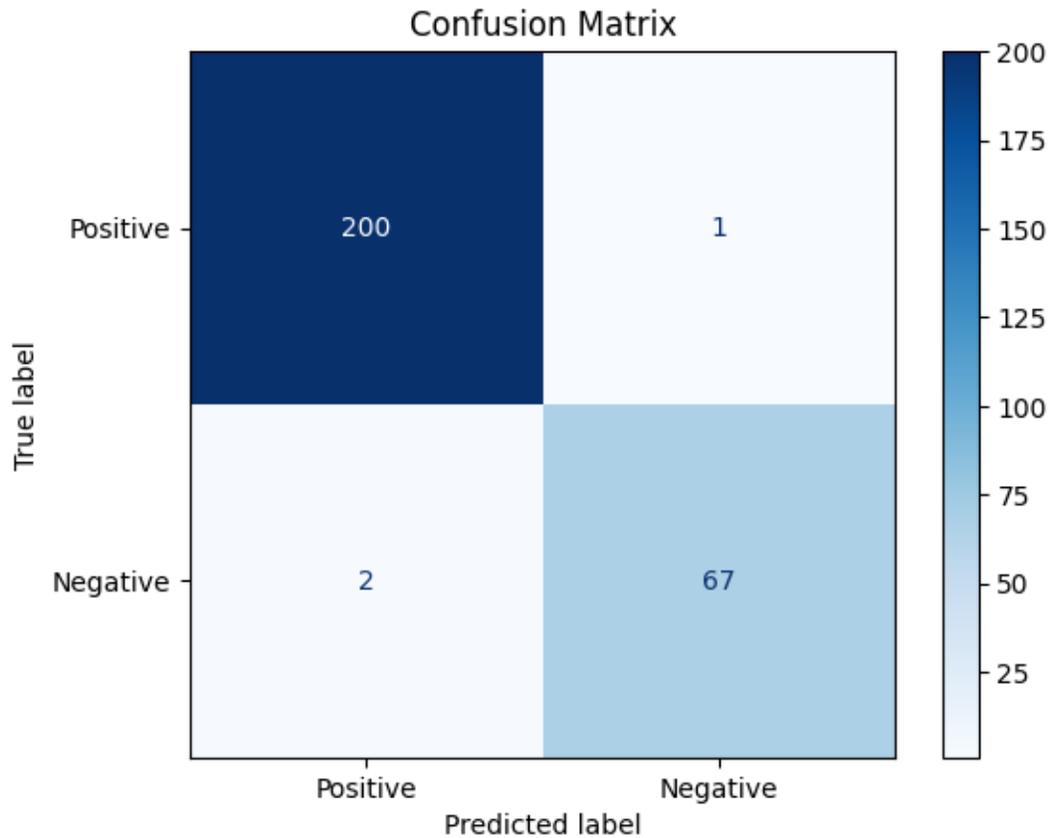

Figure 8Confusion matrix for test Data

## 5 Discussion

Brain tumor detection faces significant challenges due to the wide variability in tumor characteristics, such as size, shape, and location. Traditional image analysis approaches often struggle to accurately identify and delineate tumor regions. Our research leverages the power of domain adaptation, specifically TrAdaBoost, to enhance data quality and improve model performance. By integrating robust algorithms like Vision Transformers (ViT) and Capsule Networks (CapsNet), we refine feature extraction and attention mechanisms, ensuring that relevant features are effectively captured and semantic relationships are preserved.

Our multi-classifier architecture embodies the potential of ensemble learning in medical image analysis. By combining classifiers that employ different feature extraction and attention mechanisms, we capitalize on their unique strengths to achieve superior performance compared to individual classifiers. This approach effectively addresses the complexity of glioma tumor detection, providing a more precise and reliable diagnostic tool.

Our study highlights the effectiveness of transfer learning and domain adaptation techniques in medical imaging, particularly in the context of brain tumor detection. The significant class imbalance between tumor and non-tumor samples in both datasets underscores the importance of employing robust machine learning algorithms capable of generalizing across domains. The high accuracy achieved suggests that our model can provide valuable support to radiologists in diagnosing brain tumors, potentially leading to earlier interventions and improved patient outcomes.

The integration of diverse datasets enhances the model's robustness, allowing it to better handle variations in MRI imaging protocols and patient demographics. Future directions could explore the incorporation of additional imaging modalities or patient data to further refine the model's predictive capabilities. Additionally, investigating the interpretability of the model's predictions will be crucial for its clinical acceptance.

## 6. Conclusion

This study demonstrates the effectiveness of employing advanced machine learning techniques to enhance brain tumor detection in MRI scans. The integration of TrAdaBoost with cutting-edge algorithms, including ViT and CapsNet, significantly improves the accuracy and efficiency of our model. The promising results, obtained by leveraging dataset, highlight the feasibility and potential of automated diagnostic tools for supporting healthcare professionals in making informed decisions.

Future research will focus on refining these methodologies further and exploring their applicability across different types of brain tumors and medical imaging modalities. By advancing the state-of-the-art in brain tumor detection, we aim to facilitate earlier diagnosis and better treatment planning, ultimately improving patient outcomes in neuro-oncology.